\documentclass{emulateapj}
\usepackage{color}

\slugcomment{Accepted by the Astrophysical Journal Letters}

\shorttitle{Alignment of Outflows with Magnetic Fields} 
\shortauthors{Matsumoto, Nakazato, and Tomisaka}

\begin{document}

\title{Alignment of Outflows with Magnetic Fields in Cloud Cores}

\author{Tomoaki Matsumoto\altaffilmark{1}, 
Takeshi Nakazato\altaffilmark{2,3}, 
and 
Kohji Tomisaka\altaffilmark{4}} 

\altaffiltext{1}{Department of Humanity and Environment, Hosei
 University, Fujimi, Chiyoda-ku, Tokyo 102-8160, Japan,
matsu@i.hosei.ac.jp}
 \altaffiltext{2}{National
 Astronomical Observatory, Mitaka, Tokyo 181-8588, Japan,
nakazato@nro.nao.ac.jp}
 \altaffiltext{3}{Institut de Radioastronomie Millim\'{e}trique, 300 rue
 de la Piscine, 38406 Saint Martin d'H\`{e}res, France}
 \altaffiltext{4}{Division of Theoretical Astrophysics, National
 Astronomical Observatory, Mitaka, Tokyo 181-8588, Japan,
tomisaka@th.nao.ac.jp}

\begin{abstract}
  We estimate the polarized thermal dust emission from MHD simulations
  of protostellar collapse and outflow formation in order to
  investigate alignment of outflows with magnetic fields. The
  polarization maps indicate that alignment of an outflow with the
  magnetic field depends on the field strength inside the cloud core;
  the direction of the outflow, projected on the plane of the sky, is
  aligned preferentially with the mean polarization vector for a cloud
  core with a magnetic field strength of 80~$\mu$G, while it does not
  tend to be aligned for 50~$\mu$G as long as the 1000~AU scale is
  considered.  The direction of the magnetic field at the cloud center
  is probed by the direction of the outflow.  In addition, the
  magnetic field at the cloud center can be revealed by {\it ALMA}
  even when the source is embedded deeply in the envelope.  The
  Chandrasekhar-Fermi formula is examined using the polarization maps,
  indicating that the field strength predicted by the formula should
  be corrected by a factor of $0.24 - 0.44$.  The correction factor
  has a tendency to be lower for a cloud core with a weaker magnetic
  field.
\end{abstract}


\section{Introduction}

Magnetic fields are believed to control not only protostellar collapse,
but also formation of circumstellar disks and outflows.  Many
observations have suggested that the outflow and jet axis of the young
star is aligned preferentially along the cloud-scale magnetic field
\citep*[e.g.,][]{Cohen84,Strom86,Vrba86,Vrba88,Tamura89,Jones03}.

However, recent observations indicate a suggestion contrary to 
previous ones concerning the issue of alignment of outflows and jets with the
magnetic fields.  High-resolution observations of submillimeter
polarization have resolved the magnetic fields around young stars on a
$\sim 10^{3-4}$~AU scale, which is comparable to the outflow scale
\citep*{Momose01,Henning01,Wolf03,Vallee03}.
\citet{Wolf03} investigate alignment of outflows with
magnetic fields for Bok globules associated with Class 0 protostars
and Class I sources, suggesting that two Bok globules are associated
with outflows parallel to the magnetic fields, while the two other
globules are associated with outflows perpendicular to the magnetic
fields.  For Classical T Tauri stars (CTTSs), \citet{Menard04} estimate
orientation of the symmetry axes of the disk-jet systems in the
Taurus-Auriga region,  and
indicate that CTTSs are oriented randomly with respect to the local
magnetic field.  


Recently, \citet{Matsumoto04} (hereafter MT04) have performed MHD simulations
of the collapse of magnetized cloud cores and reproduced outflow generation
in order to investigate the directions of outflows, circumstellar
disk, rotation, and magnetic fields.  The simulations show that the
outflow tends to be aligned with a local magnetic field of a 10~AU scale
irrespective of the magnetic field strength assumed, while the
alignment depends on the field strength on the cloud core scale.
A disk-outflow system is aligned with the cloud core scale magnetic
field within $\sim 5^\circ$ and $\sim 30^\circ$ for the initial field
strengths of 37.1 and $18.6\,\mu$G, respectively, because of the
strong magnetic braking during the collapse.  When a weak field
strength of $7.42\,\mu$G is assumed, the outflow is not aligned with
the cloud core scale magnetic field.
In this Letter, alignment of an outflow with magnetic field is
discussed by constructing polarization maps from the MHD simulations
of MT04.  


\section{Cloud Model and Polarized Emission}
\label{sec:model}

Polarization of dust emission is calculated from MT04 MHD simulation
data. The simulations follow the gravitational collapse of 
cloud cores, formation of a first stellar core \citep{Larson69}, and
the launch of an outflow, resolving both the whole cloud core and the
protostar.
Polarization maps are
constructed by extracting the central cubic region (9128~AU)$^3$
from all the simulation data. 

The initial model of a cloud core is a slowly
rotating, spherical, isothermal cloud threaded by a uniform magnetic
field. The initial cloud core has the density
profile of a Bonnor-Ebert sphere \citep{Ebert1955,Bonnor1956}, and 
the present model can be applied to Bok globules because Bok globules
are thought to have Bonnor-Ebert density profiles
\citep{Alves01,Harvey01,Racca02}.
The initial central density is set as $\rho_0 =1\times10^{-19}\,{\rm
g}\,{\rm cm}^{-3}$, which corresponds to a number density of $
n_0 = 2.61 \times 10^4\,{\rm cm}^{-3}$ for the assumed mean molecular weight
of 2.3. The initial temperature of the gas is 10~K, the edge of the
cloud is located at $r =0.178$~pc from the center, 
and the mass of the cloud is $6.130\,M_\odot$. 
The initial angular velocity is assumed to be 
$7.11\times10^{-7}\,{\rm yr}^{-1}$.
In this Letter, the last stages of two models of MT04, models MF45 and WF45, are
shown.  In these models, 
the initial magnetic field
is inclined at an angle of
$\theta = 45^\circ$ from 
the initial rotation axis, which corresponds to the $z$-axis.
The initial magnetic field strengths $B_0$ are shown in 
Table~\ref{table:models}.

\begin{deluxetable}{lrrrr}
\tabletypesize{\scriptsize}
\tablecaption{Parameters and Properties of Model Cores\label{table:models}}
\tablewidth{0pt}
\tablehead{
\colhead{} & 
\colhead{$B_0$} & 
\colhead{$\bar{B}$} &
\colhead{$\bar{n}$} & 
\colhead{$\phi_\mathrm{3D}$} \\
\colhead{Models} & 
\colhead{($\mu {\rm G}$)} & 
\colhead{($\mu {\rm G}$)} &
\colhead{(cm$^{-3}$)} & 
\colhead{(deg)} 
}
\startdata
 MF45 & 18.6 &  82.8 & $2.74\times10^5$  & 12.4\\
 WF45 & 7.42 &  50.1 & $2.78\times10^5$  & 53.5
 \enddata
\end{deluxetable}

Table~\ref{table:models} also shows
the mean strength of magnetic field $\bar{B}$ and 
mean number density $\bar{n}$ within the 
region (9128~AU)$^3$ for comparison with polarization
maps in \S~\ref{sec:results}.
The mean values $\bar{B}$ and $\bar{n}$ are
 considerably larger than the initial
values $B_0$ and $n_0$ because of amplification due to collapse.
Angle $\phi_\mathrm{3D}$ 
denotes a three-dimensional angle between
the direction of the mean magnetic field 
within $r \le 50$~AU  and that within $4555$~AU,
indicating change in the direction of the magnetic field
inside the cloud core (see Figs.~7 and 15 of MT04 for models MF45
and WF45, respectively), and 
model WF45 shows an angle larger than that in 
model MF45 because of the weak
magnetic braking. 
The direction of the mean magnetic field within $r \le 50$~AU
approximately coincides with 
the local direction of the outflow,
and is adopted here as an index of the direction of the outflow.

The polarization of dust emission is calculated following the analysis
of \citet{Fiege00} and \citet{Padoan01}. As these studies do, we focus on the thermal
dust emission at submillimeter wavelengths, neglecting scattering.
The Stokes parameters $Q$ and $U$ are proportional to the following
integrals of $q$
and $u$ when we assume that the grain properties and temperature are constant:
\begin{equation}
q = \int \rho \cos 2\psi \cos^2 \gamma ds,
\end{equation}
\begin{equation}
u = \int \rho \sin 2\psi \cos^2 \gamma ds,
\end{equation}
where the integrals $\int ds $ are performed along the line of sight,
$\rho$ denotes the gas density, $\psi$ is the angle between the
projection of the magnetic field on the plane of the sky and
the north, and $\gamma$ is the angle between the plane of the sky and
the local direction of the magnetic field. The polarization angle
$\chi$ is given by
$
\tan 2\chi = u/q,
$
where $-\pi/2 \le \chi < \pi/2$.  
The polarization vector with $\chi$ traces the polarization shown by
the B-vector of the submillimeter thermal emission, which is parallel
to the interstellar magnetic field on the plane of the sky.
The degree of polarization is calculated by
$
P = \alpha \left(q^2 + u^2 \right)^{1/2} / \left(\Sigma - \alpha \Sigma_2\right)
$
with 
$
\Sigma = \int \rho ds
$
and 
$
\Sigma_2 = (1/2) \int \rho \left( \cos^2 \gamma - 2/3 \right) ds,
$
where $\alpha$ is a parameter specified by the grain properties, and we
adopt a constant value of $\alpha = 0.15$ following \citet{Padoan01}.

\begin{figure*}[t]
\epsscale{1.2}
\plotone{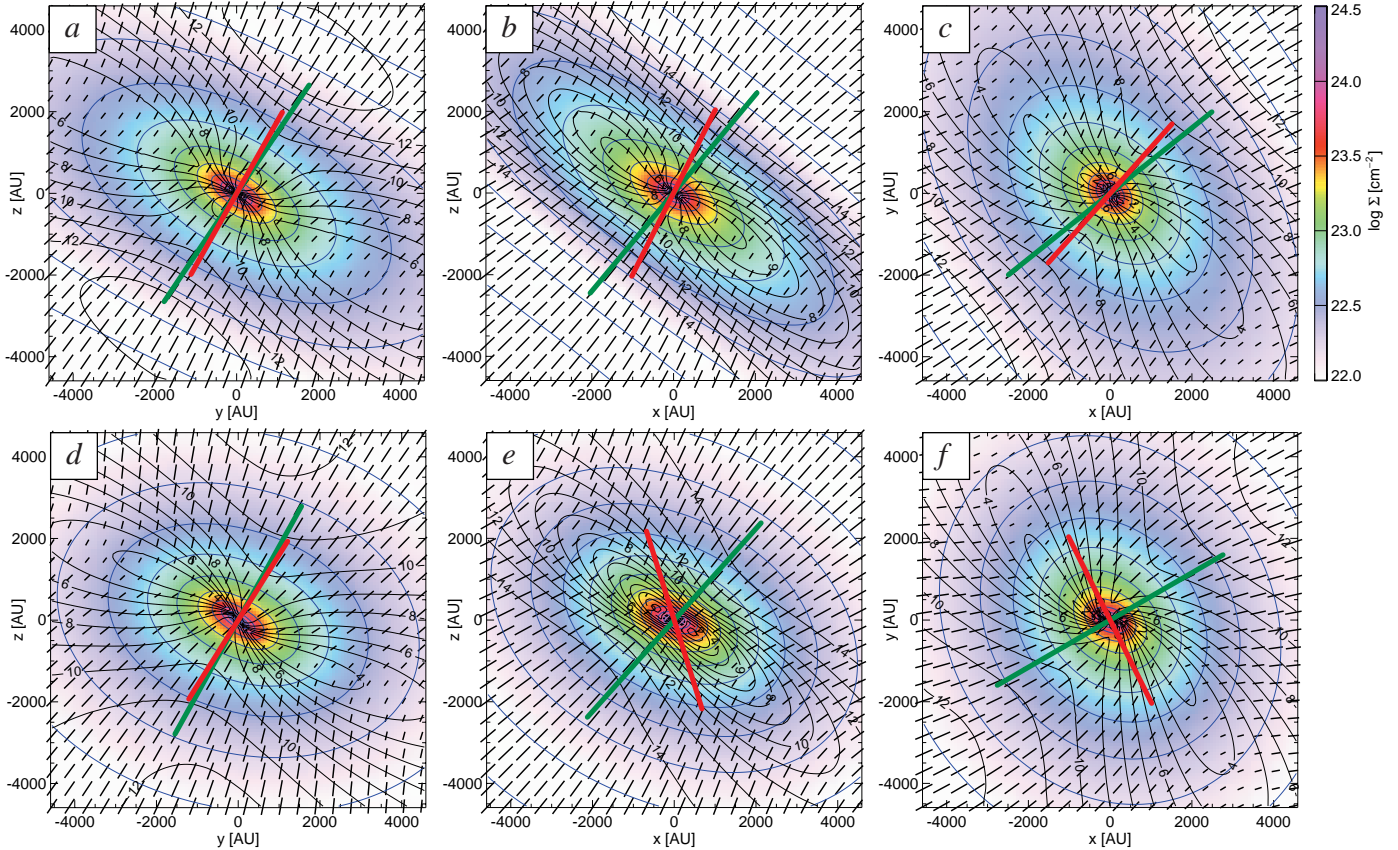}
\figcaption[f1.eps]{ 
Polarization maps constructed from MHD data of the central (9182~AU)$^3$ box for 
models ({\it a-c}) MF45 and ({\it d-f}) WF45 along the lines of sight
parallel to 
({\it a, d}) the $x$-axis,
({\it b, e}) the $y$-axis,
({\it c, f}) and the $z$-axis, respectively. 
Vector traces the B-vector of the polarized thermal emission, and
is therefore parallel to the interstellar magnetic field lines on the
plane of the sky.
The length of the vector is proportional to the 
degree of polarization, which is also shown by black contours
at 1~\% intervals.
The mean orientation of the polarization vector is denoted by
the thick green line.
Color scale and blue contours denote the surface density $\Sigma$,
which is proportional to the intensity of thermal dust emission when 
the gas is assumed to be optically thin.
Thick red line denotes the projected direction of the mean
magnetic fields averaged over $r \le  50 \mathrm{AU}$, indicating
the direction of the outflow.
\label{f1.eps}
}
\end{figure*}

\section{Results}
\label{sec:results}
\subsection{Polarization Maps of the Cloud Cores}
\label{sec:cloud}

Figure~\ref{f1.eps} shows the polarization maps overlaid by the
surface density for models MF45 and WF45 for three orthogonal lines of
sight ($x$, $y$, and $z$-directions).
The elliptical distribution of the surface density reflects
the flat infalling envelope for all the models and for
all the lines of sight (see Fig. 2 of MT04 for the
three-dimensional density and magnetic field structures of
model MF45).  
The hourglass structure of magnetic fields in three dimensions
is projected directly on the polarization vector.
The orientations of the mean polarization vector (thick green line) are
perpendicular to the long axes of the surface densities.
The degree of polarization 
along the long axis is less than that
along the short axis due to the hourglass structure of the field lines;
this field structure increases the component of the magnetic field parallel
to the line of sight $B_\mathrm{los}$ compared with the perpendicular
component $B_\perp$, and reduces the degree of polarization along the
flat envelope.

The outflows are hardly visible in the maps because the outflows are
extended up to only $\sim 200$ and 150~AU for models MF45 and WF45. 
We adopt here the direction of the mean magnetic field within $r\le50$~AU
as an index of the direction of the outflow (thick red line).
%
%
The projected direction of the outflow is almost aligned with 
the mean polarization vector (thick green line) for model MF45
 irrespective of the lines of sight (Fig.~\ref{f1.eps}{\it a-c})
because of a small intrinsic angle between the outflow and the 
cloud core scale magnetic field ($\phi_\mathrm{3D}=12.4^\circ$).
On the other hand, 
the alignment depends on 
the line-of-sight for model WF45 (Fig.~\ref{f1.eps}{\it d-f})
because the outflow is not aligned with the magnetic field
intrincically ($\phi_\mathrm{3D}=53.5^\circ$).
When model WF45 is observed along the $x$-direction
(Fig.~\ref{f1.eps}{\it d}), the projected direction of the outflow
is aligned with the mean polarization vector by chance.
They are not aligned considerably when observed along
the $y$- and $z$-directions as shown in Figure~\ref{f1.eps}{\it e} and
\ref{f1.eps}{\it f}.

\begin{deluxetable*}{lccccccc}
\tabletypesize{\scriptsize}
\tablecaption{Outflow Orientations and Estimate of Magnetic Field Strengths \label{table:pol}}
\tablewidth{0pt}
\tablehead{
\colhead{} & 
\colhead{Line of } & 
\colhead{$\phi_\mathrm{prj}$} & 
\colhead{$\bar{\rho}$} & 
\colhead{$\delta v$} & 
\colhead{$\sigma_\chi$} &
\colhead{$B_\mathrm{CF}$} &
\colhead{$f$} \\
\colhead{Models} & 
\colhead{Sight} & 
\colhead{(deg)} & 
\colhead{(g cm$^{-3}$)} & 
\colhead{(km s$^{-1}$)} &
\colhead{(deg)} & 
\colhead{($\mu$G)} &
\colhead{} 
}
\startdata
MF45 & $x$ &  4.40 & $1.04\times10^{-18}$ & 0.209 & 10.3 & 242.8 & 0.34 \\
MF45 & $y$ &  13.1 & $1.04\times10^{-18}$ & 0.214 & 11.0 & 233.2 & 0.35 \\
MF45 & $z$ &  9.62 & $1.04\times10^{-18}$ & 0.193 & 12.4 & 186.4 & 0.44 \\
WF45 & $x$ &  $-3.02$& $1.06\times10^{-18}$ & 0.262&16.0 & 197.6 & 0.25 \\
WF45 & $y$ &  59.1 & $1.06\times10^{-18}$ & 0.251 & 14.3 & 212.2 & 0.24 \\
WF45 & $z$ &  86.5 & $1.06\times10^{-18}$ & 0.244 & 17.2 & 171.6 & 0.29 
 \enddata
\end{deluxetable*}

More qualitatively,
Table~\ref{table:pol} shows $\phi_\mathrm{prj}$, which is defined as
an angle between the red and thick green lines.
This angle indicates that the direction of the outflow tends to
be aligned with the mean polarization vector on the plane of the sky
when a cloud core has a strong magnetic field of $\bar{B} \sim 80\mu$G.

The polarization maps examined here demonstrate that the direction of
the magnetic field at the cloud center can not be inferred from the
polarization on the scale of this map ($\sim 10^3$~AU scale).
Model MF45 shows polarization maps similar to model
WF45, although the magnetic field at the cloud center
of Model MF45 is projected to quite different direction from that of Model WF45.The directions of the central magnetic field reflect
the those of outflows.
In other words, the direction of the outflow probes the
direction of the magnetic field at the cloud center.

\begin{figure*}[t]
\epsscale{1.2}
\plotone{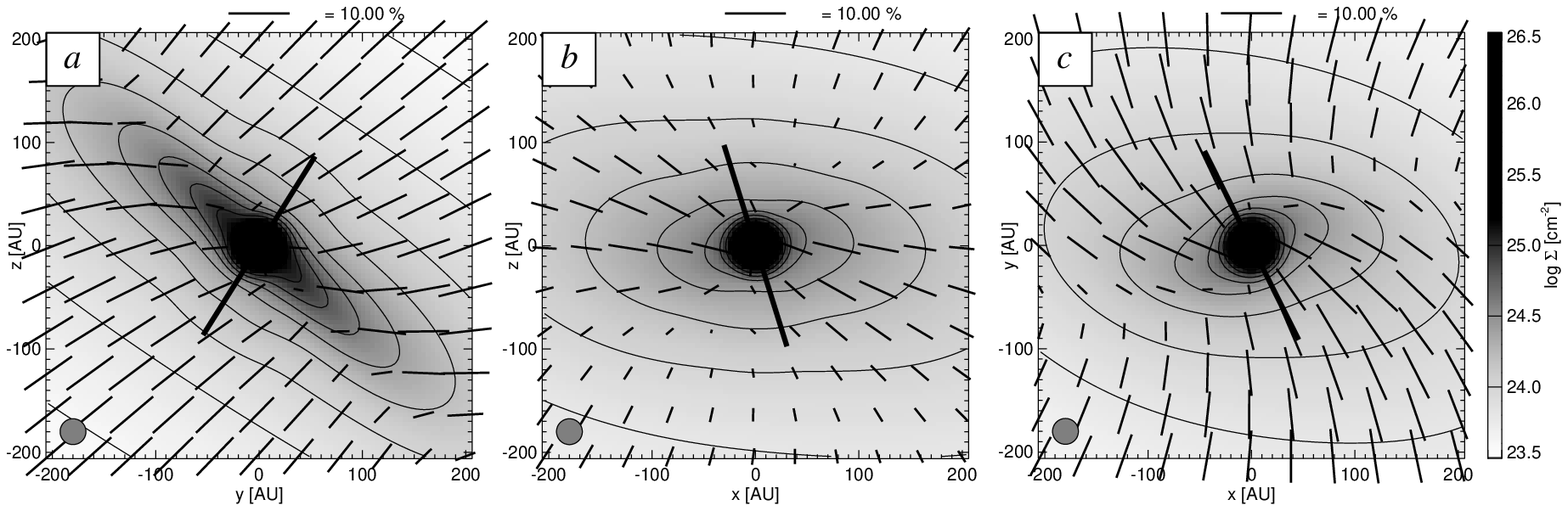}
\figcaption[f2.eps]{ 
Polarization maps convoluted by 
the Gaussian beam with $\mathrm{FWHM} = 25 \mathrm{AU}$
toward the cloud center for model WF45 
along the lines of sight
parallel to 
({\it a}) the $x$-axis,
({\it b}) the $y$-axis, and
({\it c}) the $z$-axis.
The beam pattern is denoted in the bottom left corner of each panel.
Thick line denotes
the projected direction of the mean
magnetic fields averaged over $r \le 50 \mathrm{AU}$, 
indicating the direction of the outflow.
Grayscale and contours denote the surface density.
\label{f2.eps}
}
\end{figure*}

Figure~\ref{f2.eps} simulates direct observations 
toward the cloud center on the 100~AU scale
with high resolution of 25~AU
for the three lines of sight for model WF45.
These maps are
constructed by convolution of the polarization pattern of 
Figure~\ref{f1.eps}{\it d}--\ref{f1.eps}{\it f}
with a  Gaussian beam of $\mathrm{FWHM} = 25$~AU, 
which corresponds to angular resolution of $0.1\arcsec$ for an object
at a distance of 250~pc, e.g., an observation toward B335 by {\it ALMA}.
In constructing Figure~\ref{f2.eps}, the envelope
of 9128~AU scale is taken into account similar to
Figure~\ref{f1.eps}{\it d}--\ref{f1.eps}{\it f}.
On the scale of Figure~\ref{f2.eps}, 
the polarization
reveals the convergence of alignment between the disk-outflow system
and the magnetic field; nevertheless the protostar is embedded deeply in the envelope.
Figure~\ref{f2.eps} shows the polarization depending on the
lines of sight (see also Fig.~13 of MT04 for the three-dimensional structure).
Figure~\ref{f2.eps}{\it a} shows an edge-on view of the disk as 
reproducing the considerably flat surface density, and the
polarization vector is almost perpendicular to the long axis of the
surface density.  The bipolar outflow is extended up to 150~AU vertically
on the map, although it is hardly visible in both the
polarization and the surface density. 
Figure~\ref{f2.eps}{\it b} also shows an edge-on view, exhibiting
a polarization pattern different from Figure~\ref{f2.eps}{\it a};
the polarization vector is oriented along the projected disk surface
because the radial component of the magnetic field in the hourglass structure
contributes $B_\perp$ there.
In $|z| \gtrsim 100$~AU, the polarization vector is still 
perpendicular to the disk, similar to Figure~\ref{f2.eps}{\it a}. 
Figure~\ref{f2.eps}{\it c} shows the change in the direction of the
mean magnetic field; the outflow, whose direction is indicated by
the thick line, is aligned with the central polarization vector.

\subsection{Estimate of Field Strengths from Polarization Maps}
\label{sec:B}

Magnetic field strength can be estimated from the polarization maps of
Figure~\ref{f1.eps} using the method of \citet{Chandra53}:
\begin{equation}
B_\mathrm{CF} = \left( \frac{4\pi}{3} \bar{\rho} \right)^{1/2}
\frac{\delta v}{\sigma_\chi},
\end{equation}
where 
$\bar{\rho}$ denotes the mean density, estimated by averaging the surface
density over the map,
$\delta v$ denotes the rms gas velocity, 
assumed as the velocity component parallel to the line of sight
superposed as $\delta v ^2= \int \rho v_\mathrm{los}^2 dV/\int \rho dV$,
and $\sigma_\chi$ denotes the standard deviation to the mean
orientation angle of the polarization vector.  
The derived parameters $\bar{\rho}$, $\delta v$,
$\sigma_\chi$, and the estimated magnetic field strength $B_\mathrm{CF}$
are shown in Table~\ref{table:pol} for the three orthogonal lines of sight. 

The estimated  field strengths range 
from 186 to 242~$\mu$G for model MF45, and from 171 to 212~$\mu$G for
model WF45, exhibiting significant dispersion\footnote{
We confirmed that a method proposed by \citet{Houde04} can correct
a projection effect of the magnetic field.}.
Moreover, these field strengths are
$2-4$ times larger than the mean magnetic field strengths $\bar{B}$
(see Table~\ref{table:models}), which are obtained directly from MHD data.
In other words, the prediction of the Chandrasekhar-Fermi formula should
be corrected by a factor of $f = 0.34 - 0.44$
for model MF45 and $f = 0.24 - 0.29$ for model WF45, as shown in
Table~\ref{table:pol}. 
\citet*{Ostriker01} and \citet{Padoan01} also report
such an overestimated field strength in application of 
the Chandrasekhar-Fermi formula, and 
obtain a correction factor of $f \simeq 0.4 - 0.5$, which
is consistent with our case.
\citet{Padoan01} discuss
the deviation from the prediction of the Chandrasekhar-Fermi formula as
being attributable to deviation from the linear theory, which is assumed by
\citet{Chandra53}.
The large values of $\sigma_\chi$ shown in Table~\ref{table:pol}
indicate that the Alfv\'en wave is nonlinear with large amplitude in our models.
Moreover, it is noteworthy that model WF45 exhibits larger
$\sigma_\chi$ than model MF45, 
indicating that the correction is more significant for a cloud with
weaker magnetic field.


\section{Discussion}
\label{sec:discussion}

Activity of an outflow may be discussed in terms of alignment of
the outflow with the magnetic field.
The cloud core with stronger magnetic field exhibits a faster outflow as
shown in Figure 18 of MT04, and the outflow tends to be aligned with the
polarization vector.  This indicates that the fast outflow is 
observed parallel to the magnetic field.  According to the observations
toward CTTSs, \citet{Menard04} suggest a similar tendency in
spite of the different evolutionary stage from that considered here: 
CTTSs without bright and extended outflows have a tendency to be
perpendicular to the magnetic field.

The outflows are extended up to only $150-200$~AU in the MHD simulation
data used here, while the outflows observed by molecular line emission
are extended up to a 1000~AU scale \citep[e.g.,][]{Wolf03}.
Moreover, the outflows presented here have considerably slower velocity
than that observed around young stars.
Therefore the cloud cores presented here are restricted to 
the very early evolutionary stage compared with the observed cloud cores.
This restriction arises in response to 
computational cost for solving the launch
mechanism of the outflow near the protostar.
In further stages, the magnetic field hardly seems to affect the
direction of the outflow on the scale of 1000~AU, because 
the outflow is accelerated at $r \sim 10$~AU at a speed 
comparable to the Alfv\'en velocity at this radius,
and the magnetic field strength of the envelope
decreases rapidly, proportional to $B \propto r^{-1}$.
Moreover, the protostar may be
decoupled from the magnetic field of the envelope as a result of
efficient ambipolar 
diffusion \citep*{Nakano02} in further stages. The directions of the
outflow will be fixed during the main accretion phase of the protostar.

\acknowledgments

Numerical computations were carried out on the VPP5000 supercomputer at the
Astronomical Data Analysis Center (ADAC) of the National Astronomical
Observatory, Japan.
This research was supported in part by
Grants-in-Aid 
for Young Scientists (B) 16740115 (TM), 
for Scientific Research (C) 14540233 (KT) and 17540212 (TM), 
and for Scientific Research (B) 17340059 (TM, KT) 
by the Ministry of Education, Culture, Sports, Science and Technology, Japan.

\end{document}